# Invited Discussion

Merlise A. Clyde*

## 1 Introduction

The article by García-Donato and co-authors addresses the dual challenges of accounting for model uncertainty and missing data within the Gaussian regression frameworks from an objective Bayesian perspective. Through the use of an imputation $g$-prior that replaces $\mathbf{X}_\gamma^T\mathbf{X}_\gamma$ in the covariance of $\boldsymbol{\beta}_\gamma$ with $\boldsymbol{\Sigma}_{\mathbf{X}_\gamma}$, the authors develop a coherent approach to addressing the missing data problem and model uncertainty simultaneously with random $\mathbf{X}_\gamma$ in the missing at random (MAR) or missing completely at random (MCAR) settings, while still being computationally tractable.

Under the MCAR/MAR framework and assumptions on the prior distributions, one of the key results that permits tractable computation is the expression for the marginal distribution of $\mathbf{Y}_{\mathsf{obs}}$ given $\mathbf{X}_{\mathsf{obs}}$ and $\boldsymbol{\gamma}$

$$m_{\boldsymbol{\gamma}}(\mathbf{Y}_{\mathsf{obs}} \mid \mathbf{X}_{\mathsf{obs}}) = \int \left[ \int f(\mathbf{Y}_{\mathsf{obs}} \mid \mathbf{X}_{\mathsf{obs}}, \mathbf{X}_{\mathsf{miss}}, \alpha, \boldsymbol{\beta}_{\boldsymbol{\gamma}}, \sigma^2, \boldsymbol{\gamma}) \pi(\alpha, \boldsymbol{\beta}_{\boldsymbol{\gamma}}, \sigma^2 \mid \boldsymbol{\nu}, \boldsymbol{\gamma}) d[\alpha, \boldsymbol{\beta}_{\boldsymbol{\gamma}}, \sigma^2] \right]$$
$$\times p(\mathbf{X}_{\mathsf{miss}} \mid \mathbf{X}_{\mathsf{obs}}, \boldsymbol{\nu}) \pi(\boldsymbol{\nu} \mid \mathbf{X}_{\mathsf{obs}}) d[\mathbf{X}_{\mathsf{miss}}, \boldsymbol{\nu}]$$

where $\boldsymbol{\nu} = (\boldsymbol{\mu}, \boldsymbol{\Sigma})$ are the hyper-parameters of the distribution of $\mathbf{X}$. By replacing the usual $\frac{1}{n}(\mathbf{X}_\gamma - \mathbf{1}_n \bar{\mathbf{x}}_\gamma)^T(\mathbf{X}_\gamma - \mathbf{1}_n \bar{\mathbf{x}}_\gamma)$ in the $g$-prior with its expectation, $\boldsymbol{\Sigma}_{\gamma\gamma}$, the inner integral (the marginal likelihood under the complete data $(\mathbf{Y}_{\mathsf{obs}}, \mathbf{X}_{\mathsf{obs}}, \mathbf{X}_{\mathsf{miss}})$ conditional on $\boldsymbol{\nu}$) continues to be available in closed form due to conjugacy. The outer integral over $\mathbf{X}_{\mathsf{miss}}$ and $\boldsymbol{\nu}$ is not tractable, but through the factorization of the posteriors, can be approximated via Monte Carlo integration leading to a "Rao-Blackwellized" estimator.

$$\boldsymbol{\nu}^{(j)} \sim p(\boldsymbol{\nu} \mid \mathbf{X}_{\mathsf{obs}})$$
$$\mathbf{X}_{\mathsf{miss}}^{(j)} \sim p(\mathbf{X}_{\mathsf{miss}} \mid \mathbf{X}_{\mathsf{obs}}, \boldsymbol{\nu}^{(j)})$$
$$m(\mathbf{Y}_{\mathsf{obs}} \mid \mathbf{X}_{\mathsf{obs}}, \mathbf{X}_{\mathsf{miss}}^{(j)}, \boldsymbol{\gamma}, \boldsymbol{\nu}^{(j)}) = \int f(\mathbf{Y}_{\mathsf{obs}} \mid \mathbf{X}_{\mathsf{obs}}, \mathbf{X}_{\mathsf{miss}}^{(j)}, \alpha, \boldsymbol{\beta}_{\boldsymbol{\gamma}}, \sigma^2, \boldsymbol{\gamma})$$
$$\times \pi(\alpha, \boldsymbol{\beta}_{\boldsymbol{\gamma}}, \sigma^2 \mid \boldsymbol{\nu}^{(j)}, \boldsymbol{\gamma}) d[\alpha, \boldsymbol{\beta}_{\boldsymbol{\gamma}}, \sigma^2]$$
$$\hat{m}_{\boldsymbol{\gamma}}(\mathbf{Y}_{\mathsf{obs}} \mid \mathbf{X}_{\mathsf{obs}}) = \frac{1}{J} \sum_{j=1}^{J} m(\mathbf{Y}_{\mathsf{obs}} \mid \mathbf{X}_{\mathsf{obs}}, \mathbf{X}_{\mathsf{miss}}^{(j)}, \boldsymbol{\gamma}, \boldsymbol{\nu}^{(j)}).$$

This is similar in spirit to the Impute Then Select (ITS) approach of Yang et al. (2005) although rather than using the same set of imputed $\mathbf{X}_{\mathsf{miss}}, \boldsymbol{\nu}$ across all models, it appears that here new imputations are generated for each model – I would expect that using the same set of imputations across all models would reduce the Monte Carlo error for estimating posterior model probabilities. Beyond the substantial difference in choice of prior distributions, the implementation of ITS in Yang et al. (2005) does not take advantage of collapsing the Gibbs sampler over $\alpha$ and $\boldsymbol{\beta}_{\boldsymbol{\gamma}}$, which would lead to less efficient estimators Liu et al. (1995); Ghosh and Clyde (2011).

## 2 Imputation *g*-Prior

For those who oppose *g*-priors on philosophical grounds because of the dependence on the design matrix $\mathbf{X}_{\boldsymbol{\gamma}}$ and potentially sample size $n$, the imputation *g*-prior may be more palatable as it replaces the sample covariance of $\mathbf{X}_{\boldsymbol{\gamma}}$ with the population parameter $\boldsymbol{\Sigma}_{\boldsymbol{\gamma}\boldsymbol{\gamma}}$. Under random sampling this is the limiting version of the *g*-prior with $g = n$ as $n \to \infty$, satisfying the fourth criterion "Intrinsic Prior Consistency" of Bayarri et al. (2012).

The imputation *g*-prior of course could be used as the "real prior" with complete data. The expression for the Bayes factor in (24) is still valid with no missing observations based on the hierarchical prior, however, there is no need to integrate over $\mathbf{X}_{\mathsf{miss}}$ in this case. Because of the non-linearity in $\boldsymbol{\Sigma}$, it was not clear to me that with complete data, the imputation prior and usual *g*-prior would lead to the equivalent results in finite samples, as there is uncertainty about $\boldsymbol{\Sigma}$ that should be propagated. There are computational advantages, however, of the usual *g*-prior over the imputation *g*-prior with complete data, which may limit its adoption in general even with random sampling. An alternative approach is to continue using the usual *g*-prior based on the observed $\mathbf{X}_{\mathsf{obs}}$ and imputed $\mathbf{X}_{\mathsf{miss}}$ instead of the imputation *g*-prior with or without missing data. This would provide an avenue to handle transformations of $\mathbf{X}$ such as quadratic terms and interactions. For example, in many analyses of the ozone data, linear, quadratic, and two-way interactions are considered, e.g. (Casella and Moreno, 2006; Liang et al., 2008). While joint normality may be appropriate for the linear terms, joint normality is not appropriate for all variables in the design matrix. While this might preclude use of the imputation *g*-prior, one could proceed with the imputation step as described for the missing linear terms, and then use the completed $\mathbf{X}$ to construct the usual *g*-prior for each model. A related approach is described in the context of multiple imputation of missing covariates with non-linear and interactions by Seaman et al. (2012) who compare 'passive imputation' of the non-linear and interaction terms with passive imputation with predictive matching (PMM), to 'just another variable' (JAV) imputation where the non-linear and interaction terms are treated as additional variables in the multivariate imputation model. When $\mathbf{X}$ is missing at random, JAV may be biased, but this bias was generally less than for passive imputation and PMM. When quadratic effects were pronounced, they found that JAV sometimes led to large bias and poor coverage. For logistic regression, JAV's performance was sometimes very poor, with PMM generally improving on passive imputation, in terms of bias and coverage, but did not eliminate the bias. Clearly more work is needed in this area to extend Bayesian Variable Selection (BVS) and Bayesian Model Averaging methods to handle non-linearity and interactions with missing data.

## 3 Scalability to Large Model Spaces

Estimates of posterior model probabilities

$$\hat{p}(\boldsymbol{\gamma} \mid \mathbf{Y}_{\mathsf{obs}}, \mathbf{X}_{\mathsf{obs}}) = \frac{\hat{m}_{\boldsymbol{\gamma}}(\mathbf{Y}_{\mathsf{obs}} \mid \mathbf{X}_{\mathsf{obs}})\pi(\boldsymbol{\gamma})}{\sum_{\boldsymbol{\gamma} \in \mathbf{G}} \hat{m}_{\boldsymbol{\gamma}}(\mathbf{Y}_{\mathsf{obs}} \mid \mathbf{X}_{\mathsf{obs}})\pi(\boldsymbol{\gamma})} \quad (1)$$

and marginal posterior inclusion probabilities are feasible for model spaces where enumeration is possible. However, for larger model spaces that preclude enumeration, Markov chain Monte Carlo (MCMC) of some form will be necessary to sample models with estimates of marginal posterior probabilities and inclusion probabilities based on ergodic averages of $\boldsymbol{\gamma}$. While one could iterate over $J$ imputations of $\mathbf{X}_{mis}$ and $\boldsymbol{\nu}$ for each model to obtain more accurate estimates of the marginal likelihoods and Bayes factors used to accept proposals of $\gamma$, using the Monte Carlo averaged marginal likelihoods and summing over sampled models will lead to biased estimates as the normalizing constant in (1) is underestimated (Clyde and Ghosh, 2012). As with ITS, one could use a two stage approach, by imputing $J$ sets of $\mathbf{X}_{\mathsf{miss}}$ and $\boldsymbol{\nu}$ and then running a standard MCMC for $M$ iterations to sample models given each imputed dataset, which has the advantage of being parallelizable. Alternatively, one could embed the imputation step within the MCMC similar to the Simultaneous Impute and Select (SIAS) algorithm of Yang et al. (2005), but instead draw new imputations of $\mathbf{X}_{\mathsf{miss}}$ and $\boldsymbol{\nu}$ given $\mathbf{X}_{\mathsf{obs}}$ at each iteration and accepting or rejecting the proposed $\boldsymbol{\gamma}, \mathbf{X}_{\mathsf{miss}}, \boldsymbol{\nu}$, where the imputation step requires sampling $\boldsymbol{\nu} = (\boldsymbol{\mu}, \boldsymbol{\Sigma})$, given $\mathbf{X}_{\mathsf{obs}}$ and generating $\mathbf{X}_{\mathsf{miss}}$ conditional on $\boldsymbol{\mu}$ and $\boldsymbol{\Sigma}$ as before. If one is estimating posterior probabilities based on ergodic averages from a MCMC, a question that arises is whether this is more efficient than proposing $\mathbf{X}_{\mathsf{miss}}$ given $\mathbf{Y}_{\mathsf{obs}}, \mathbf{X}_{\mathsf{obs}}$ and $\boldsymbol{\nu}$ in a Gibbs step and then accepting/rejecting a proposed $\boldsymbol{\gamma}$ given the completed $\mathbf{X}$ and $\mathbf{Y}_{\mathsf{obs}}$ in a Metropolis Hastings step. Using the information in $\mathbf{Y}_{\mathsf{obs}}$ may lead to more informative proposals for the missing data, at the cost of potentially slower mixing.

### 3.1 Objective Graphical Model Selection and *g*-Priors

For large $p$, sampling the full $p \times p$ covariance matrix $\boldsymbol{\Sigma}$ may be memory intensive and computationally expensive depending on the rejection rate for proposing $\boldsymbol{\Sigma}$; a concern if $\boldsymbol{\Sigma}$ is nearly singular (Sun and Berger, 2007). One potential approach to reduce the dimension of $\boldsymbol{\Sigma}$ is to consider selection in the covariance structure of $\mathbf{X}$ in addition to the variables in the regression of $\mathbf{Y}$ on $\mathbf{X}$. We define the vector

$$\mathbf{Z} = \begin{pmatrix} Y \\ \mathbf{x}_{\boldsymbol{\gamma}} \\ \mathbf{x}_{-\boldsymbol{\gamma}} \end{pmatrix} \in \mathbb{R}^{p+1}$$

where $Y$ is a scalar, $\mathbf{x}_{\boldsymbol{\gamma}}$ are the elements of $\mathbf{x}$ where $\gamma_j = 1$ and $\mathbf{x}_{-\boldsymbol{\gamma}}$ are the elements of $\mathbf{x}$ where $\gamma_j = 0$. Then $\mathbf{Z}$ has a multivariate normal distribution implied by the conditional distribution of $Y \mid \mathbf{x}_{\boldsymbol{\gamma}}$ and the marginal distribution of $\mathbf{x}$,

$$\mathbf{Z} =\mid \boldsymbol{\gamma}, \boldsymbol{\mu}, \boldsymbol{\Sigma} \sim \mathsf{N}\left(\begin{pmatrix} \mu_y \\ \mu_{\mathbf{x}_{\boldsymbol{\gamma}}} \\ \mu_{\mathbf{x}_{-\boldsymbol{\gamma}}} \end{pmatrix}, \boldsymbol{\Sigma} = \begin{pmatrix} \Sigma_{yy} & \boldsymbol{\Sigma}_{y\mathbf{x}_{\boldsymbol{\gamma}}} & \boldsymbol{\Sigma}_{y\mathbf{x}_{-\boldsymbol{\gamma}}} \\ \boldsymbol{\Sigma}_{\mathbf{x}_{\boldsymbol{\gamma}}y} & \boldsymbol{\Sigma}_{\mathbf{x}_{\boldsymbol{\gamma}}\mathbf{x}_{\boldsymbol{\gamma}}} & \boldsymbol{\Sigma}_{\mathbf{x}_{\boldsymbol{\gamma}},\mathbf{x}_{-\boldsymbol{\gamma}}} \\ \boldsymbol{\Sigma}_{\mathbf{x}_{-\boldsymbol{\gamma}}y} & \boldsymbol{\Sigma}_{\mathbf{x}_{-\boldsymbol{\gamma}}\mathbf{x}_{\boldsymbol{\gamma}}} & \boldsymbol{\Sigma}_{\mathbf{x}_{-\boldsymbol{\gamma}}\mathbf{x}_{-\boldsymbol{\gamma}}} \end{pmatrix}\right)$$

with conditional distribution of $Y \mid \mathbf{x_\gamma}$ given by

$$Y \mid \mathbf{x_\gamma}, \boldsymbol{\gamma} \sim \mathsf{N}(\mu_{Y|\mathbf{x_\gamma}}, \sigma^2_{y|\mathbf{x_\gamma}})$$
$$\mu_{Y|\mathbf{x_\gamma}} = \mu_y + \boldsymbol{\Sigma}_{y\mathbf{x_\gamma}} \boldsymbol{\Sigma}^{-1}_{\mathbf{x_\gamma x_\gamma}} (\mathbf{x_\gamma} - \mu_{\mathbf{x_\gamma}})$$
$$\sigma^2_{y|\mathbf{x_\gamma}} = \Sigma_{yy} - \boldsymbol{\Sigma}_{y\mathbf{x_\gamma}} \boldsymbol{\Sigma}^{-1}_{\mathbf{x_\gamma x_\gamma}} \boldsymbol{\Sigma}_{\mathbf{x_\gamma} y}$$

and marginal distribution of $\mathbf{x}$ given by

$$\begin{pmatrix} \mathbf{x_\gamma} \\ \mathbf{x_{-\gamma}} \end{pmatrix} \sim \mathsf{N}\left( \begin{pmatrix} \boldsymbol{\mu}_{\mathbf{x_\gamma}} \\ \boldsymbol{\mu}_{xng} \end{pmatrix}, \begin{pmatrix} \boldsymbol{\Sigma}_{\mathbf{x_\gamma x_\gamma}} & \boldsymbol{\Sigma}_{\mathbf{x_\gamma x_{-\gamma}}} \\ \boldsymbol{\Sigma}_{\mathbf{x_{-\gamma} x_\gamma}} & \boldsymbol{\Sigma}_{\mathbf{x_{-\gamma} x_{-\gamma}}} \end{pmatrix} \right).$$

Under model $\boldsymbol{\gamma}$, $\Sigma_{Y\mathbf{x_{-\gamma}}} = \mathbf{0}$. While $\sigma^2_{y|\mathbf{x_\gamma}}$ is generally treated as constant across models, it does appear to depend on the model through $\boldsymbol{\Sigma}_{\mathbf{x_\gamma x_\gamma}}$ and $\Sigma_{y\mathbf{x_\gamma}}$.

This may be viewed as a special case of model selection in Gaussian graphical models, where the graph $\mathbf{G}$ is defined by edges between all pairs of variables in $\mathbf{x}$ (a fully connected graph) and edges between $Y$ and variables in $\mathbf{x_\gamma}$. By relaxing, the assumption that $\mathbf{x}$ is fully connected, one could consider more general graphical models for $\mathbf{Z}$ leading to more parsimonious models for large $p$ (Jones et al., 2005). As with model specific parameters in linear regression, neither improper or vague priors on $\boldsymbol{\Sigma}$ are generally allowable. In all but the smallest of problems, $\pi(\boldsymbol{\Sigma} \mid \mathbf{G})$ must be a conjugate hyper-inverse Wishart prior (HIW) (Dawid and Lauritzen, 1993; Giudici and Green, 1999) for tractable computation of marginal likelihoods. Seeking well behaved objective priors for graphical model selection, Carvalho and Scott (2009) developed a HIW g-prior, $\boldsymbol{\Sigma} \sim \mathsf{HIW}_{\mathbf{G}}(gn, g\mathbf{Z}^T\mathbf{Z})$ constructed using fractional Bayes factors (O'Hagan, 1995). Under this approach, some fraction, $0 < g < 1$ of the likelihood is used for training an improper prior on $\boldsymbol{\Sigma}$, where the resulting fractional prior is proportional to the improper prior times the fractional likelihood. Integrating the remaining $1 - g$ fraction of the likelihood with respect to the fractional prior leads to a fractional marginal likelihood. They recommend the choice $g = 1/n$ based on minimal training samples leading to the vector $\mathbf{z}$ having a marginal Cauchy distribution and prove that the resulting Bayes Factors avoid the information paradox (Liang et al., 2008; Bayarri et al., 2012).

Focusing on the implied conditional regression for $\mathbf{Y} \mid \mathbf{X_\gamma}$ implied by the graph Carvalho and Scott (2009) prove the fractional prior for $\boldsymbol{\Sigma}$ induces a *g*-prior for $\boldsymbol{\beta_\gamma} = \Sigma_{y\mathbf{x_\gamma}} \boldsymbol{\Sigma}^{-1}_{\mathbf{x_\gamma x_\gamma}}$,

$$\boldsymbol{\beta_\gamma} \mid \sigma^2_{y|\mathbf{x_\gamma}} \sim \mathsf{N}\left( \hat{\boldsymbol{\beta}}_{\boldsymbol{\gamma}}, \frac{\sigma^2_{y|\mathbf{x_\gamma}}}{g} (\mathbf{X}'_{\boldsymbol{\gamma}} \mathbf{X}_{\boldsymbol{\gamma}})^{-1} \right) \qquad (2)$$

$$1/\sigma^2_{y|\mathbf{x_\gamma}} \sim \mathsf{Ga}\left( \frac{gn + p_{\boldsymbol{\gamma}}}{2}, \frac{g\mathsf{RSS}_{\mathbf{x_\gamma}}}{2} \right) \qquad (3)$$

where $\hat{\boldsymbol{\beta}}_{\boldsymbol{\gamma}}$ is the usual least squares estimate and $\mathsf{RSS}_{\mathbf{x_\gamma}}$ is the residual sum of squares for regression of $\mathbf{Y}$ on $\mathbf{X_\gamma}$ and $p_{\boldsymbol{\gamma}} = \sum \gamma_j$. In the case where no selection in the covariance structure of $\mathbf{X}$ is done, this provides an alternative justification of a *g*-prior using the completed $\mathbf{X}$.

The general problem of objective graphical model selection with the HIW *g*-prior is with missing data is an interesting direction to pursue. One potential approach under the HIW *g*-prior is to initiate the chain at some draw of $\mathbf{Z}_{\mathsf{miss}}^{(0)}$ and iterate for $i = 1, \ldots, M$:

Propose $\mathbf{G}^* \sim q(\mathbf{G}* \mid \mathbf{G}^{(i)})$

Propose $\boldsymbol{\mu}^*, \boldsymbol{\Sigma}^* \mid \mathbf{G}^*, \mathbf{Z}_{\mathsf{miss}}^{(i)}, \mathbf{Z}_{\mathsf{obs}}$ from the joint posterior conditional

Propose $\mathbf{Z}_{\mathsf{miss}}^* \mid \mathbf{Z}_{\mathsf{obs}}, \boldsymbol{\mu}^*, \boldsymbol{\Sigma}^*, \mathbf{G}^*$ from the posterior conditional

Accept/Reject $\mathbf{G}^*, \boldsymbol{\mu}^*, \boldsymbol{\Sigma}^*, \mathbf{Z}_{\mathsf{miss}}^*$ based on the standard Metropolis Hastings ratio.

Update $\mathbf{G}^{(i+1)}, \boldsymbol{\mu}^{(i+1)}, \boldsymbol{\Sigma}^{(i+1)}, \mathbf{Z}_{\mathsf{miss}}^{(i+1)}$

at iteration $i+1$ to either the proposed values (Accept) or the previous values (Reject). For complete data, we can collapse by integrating out $(\boldsymbol{\mu}, \boldsymbol{\Sigma})$. Other factorizations, approaches to collapse the conditional distributions or approximations for proposing $\mathbf{Z}_{\mathsf{miss}}$ given $\mathbf{G}$ could lead to more efficient sampling approaches that improve mixing. The choice of improper prior in Carvalho and Scott (2009) leads to a computationally convenient fractional prior as a *g*-prior for graphical models. Further improvements, could be achieved by adapting recent recommendations of Berger et al. (2020) for objective priors on $\boldsymbol{\Sigma}$.